\documentclass[a4paper,11pt]{article}
\usepackage{pos}
\usepackage{wrapfig}

\title{The trigger design for AdvCam}

\author*[a]{Leonid Burmistrov}
\author[a]{Matthieu Heller}
\author[a]{Luca Giangrande}
\author[a]{Tjark Miener}
\author[a]{Teresa Montaruli}
\author[d]{Iaroslava Bezshyiko}
\author[d]{Carlos Abellan Beteta}

\affiliation[a]{Département de Physique Nucléaire et Corpusculaire, Université de Genève, Faculté de Sciences,\\
1211 Genève 4, Switzerland}

\affiliation[d]{Physik-Institute of the University of Zurich, Switzerland}

\author[b]{Jorge Buces}
\author[b]{Dafne Martín-Domínguez}
\author[b]{Daniel Nieto}
\author[b]{María Molina Delicado}
\author[b]{Alejandro Pérez Aguilera}
\author[b]{Juan Abel Barrio}
\author[b]{Luis Ángel Tejedor Álvarez}

\affiliation[b]{Institute of Particle and Cosmos Physics, Universidad Complutense de Madrid,\\
Pza. Ciencias 1, Madrid E-28040, Spain}

\author[c]{Andres Upegui Posada}
\author[c]{Tanguy Dietrich}
\author[c]{Quentin Berthet}

\affiliation[c]{Haute École du Paysage, d'Ingénierie et d'Architecture \\
Rue de la Prairie 4 CH-1202 Genève, Switzerland}
 
\emailAdd{leonid.burmistrov@unige.ch}

\abstract{The AdvCam is a next-generation camera for the Large-Sized Telescopes of the Cherenkov Telescope Array Observatory, based on silicon photomultipliers. Its fully digital readout system enables the design of new, sophisticated trigger logic.

The Large-Sized Telescopes aim to cover the low-energy range of the cosmic gamma-ray spectrum, with a threshold starting at about 20~GeV, using the existing photomultiplier tube camera. The AdvCam, along with the new trigger logic, as shown by simulations, lowers the detectable energy threshold to 13~GeV.

The proposed trigger logic has a multilevel structure. The first level involves fast coincidences among small pixel regions at a rate of approximately 1~GHz, while the second level processes all camera pixels within an approximately 10-nanosecond time window. Different families of machine learning algorithms optimized for FPGAs form the second-level trigger. In this work, we consider two main approaches: Deep Neural Networks and Density-Based Spatial Clustering of Applications with Noise, both running with latencies below 1 microsecond at a 1 MHz rate. This work provides a detailed description of the trigger chain and its performance, as studied through simulation.}

\ConferenceLogo{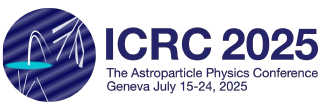}

\FullConference{39th International Cosmic Ray Conference (ICRC2025)\\
15–24 July 2025\\
Geneva, Switzerland\\}

\begin{document}
\maketitle

\section{Introduction}
The Cherenkov Telescope Array Observatory (CTAO)~\cite{ACHARYA20133} is the next-generation ground-based facility designed to detect high-energy gamma rays. It uses an array of telescopes to capture the Cherenkov light produced when gamma rays interact with Earth's atmosphere, as described in more detail~\cite{hillas2013evolution}. CTAO will provide unprecedented sensitivity and precision in energy and direction reconstruction, helping to study extreme cosmic phenomena such as supernovae,  neutron stars, black holes. CTAO is a powerful instruments for studying dark matter by searching for gamma-ray signals resulting from its annihilation or decay. 

The CTAO will be built on two sites, one in the Northern hemisphere (La Palma, Spain) and one in the Southern hemisphere (Atacama Desert, Chile) to provide full coverage of the sky. The observatory consists of three types of telescopes, each optimized to detect gamma rays of different energies. \textbf{LSTs:} Large-Sized Telescopes, with a mirror diameter of 23~m, a low-energy threshold of about $\sim$20~GeV, and a field of view of 4.5$^{\circ}$. \textbf{MSTs:} Medium-Sized Telescopes, with a mirror diameter of 12~m, a low-energy threshold of about $\sim$100~GeV, and a field of view of 8$^{\circ}$. \textbf{SSTs:} Small-Sized Telescopes, with a mirror diameter of 4~m, a low-energy threshold of about $\sim$1~TeV, and a field of view of $\sim 9^{\circ}$.

\section{SiPM camera}

Reducing the energy threshold requires detecting a greater amount of Cherenkov light. Thus going for larger mirror dishes would decrease the energy threshold. However, the LST reaches the limits of mechanical lightness and agility required for rapid repositioning, while keeping the structure affordable to build. An alternative approach is to use light sensors with improved photo-detection efficiency (PDE). Modern Silicon Photomultiplier (SiPM) achieve a maximum PDE of about 70~$\%$, which surpasses the theoretical limit of classical Photomultiplier Tubes (PMT). Additionally, the crosstalk is reduced to less than ~5$\%$ while maintaining a PDE of around 70$\%$~\cite{nemallapudi2024nuv}. SiPMs are already used in Cherenkov telescopes, such as SST-1M~\cite{heller2017innovative}, Terzina~\cite{instruments8010013,Burmistrov_2023}. 

\begin{figure}[t]
    \centering
    \includegraphics[width=15cm]{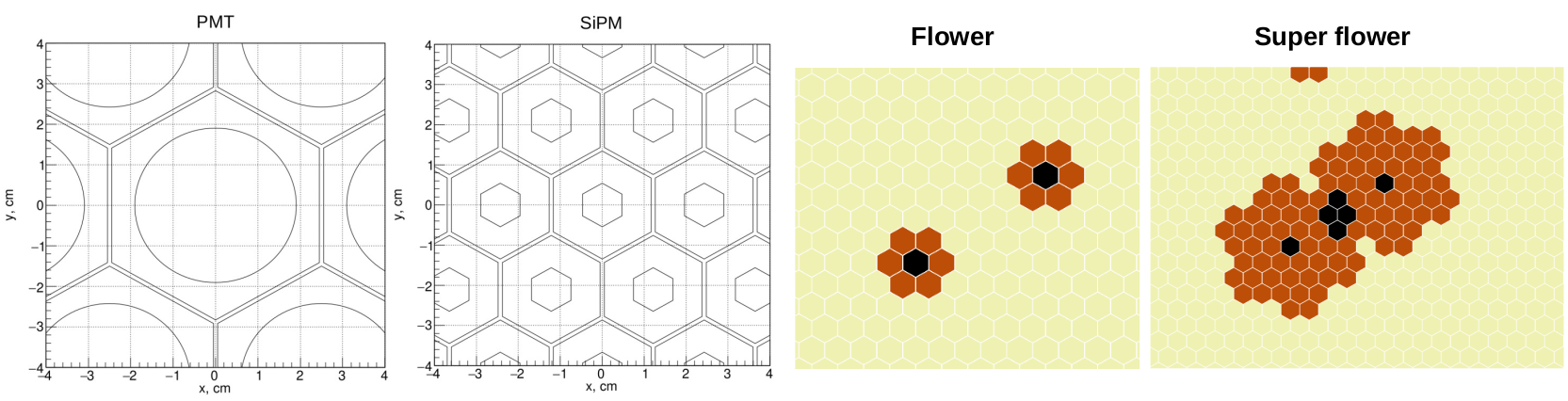}
    \caption{(\textbf{Left}): A close-up view of the PMT and SiPM-based cameras. The PMTs are round and have a diameter of $\sim 4$~cm. The hexagonal outline around it shows the light guide interfaced to the PMT. The SiPMs and their light guides have a hexagonal shape. The flat-to-flat distance of the SiPM is approximately 2.2~cm. (\textbf{Right}): Pixel grouping patterns referred to as : flower and super-flower. The flower pattern consists of a central pixel, or seed (shown in black), with 6 surrounding pixels (total of 7 channels). The super-flower pattern consists of a central flower surrounded by 6 flowers (total of 49 channels).}
    \label{fig:PMT_vs_SiPM}
\end{figure}

These factors motivate the AdvCam project~\cite{heller2023next}, which aims to develop a SiPM-based camera for LST. The camera has four times more pixels (compared to current PMT camera) for the same field of view enabling finer image resolution~(Figure~\ref{fig:PMT_vs_SiPM}, left panel). It is important to mention that the main challenge is that the increased PDE of SiPMs also increases their sensitivity to night sky background (NSB). Therefore, new triggering and filtering algorithms play an important role in enhancing sensitivity. 


\section{Simulation pipeline}
Our study is based on Monte Carlo simulations of protons, gammas, and electrons. Their interactions with the atmosphere, including the initiation, development of extensive air showers and the propagation of Cherenkov light through the atmosphere, are simulated by Corsika~7.7~\cite{heck1998corsika}. Ray-tracing, optical simulation of the telescope, and the electronics response are simulated using sim\_telarray (2020-06-29 ver.)~\cite{bernlohr2008simulation}. The NSB level is set to 268~MHz per AdvCam channel. We consider a 14-bit ADC with a 1.024~GHz waveform digitization frequency, 8.25~ADC counts per photoelectron (p.e.), and 3.94 ADC counts of electronic noise (RMS). The recorded waveform length is 75 points, corresponding to a $\sim$73.2~ns time window.

\begin{table}
\centering
\begin{tabular}{lcccc}
\hline
 & \textbf{Gamma on axis} & \textbf{Gamma diffuse} & \textbf{Proton} & \textbf{Electron} \\
\hline
\textbf{Energy, TeV}                 & 0.005 - 50 & 0.005 - 50 & 0.01 - 100 & 0.005 - 5 \\
\textbf{View cone, degree}           & 0          & 6          & 10         & 6         \\
\textbf{Generation circle radius, m} & 800        & 1000       & 1500       & 1000      \\
\hline
\end{tabular}
\caption{Simulation parameters for different particle types. Each shower is reused 10 times with different impact parameters.}
\label{tab:simulation_params}
\end{table}



We performed trigger-less simulations for four LSTs, requiring only a minimum number of p.e. to be detected by a single telescope, and considered a single pointing direction (20$^{\circ}$, zenith and 180$^{\circ}$, azimuth). The energy spectrum we simulate follows a $1/\mathrm{E}^{2}$ distribution for each particle type. For proton rate estimation, we use two smoothly broken power-law models, taken from the DAMPE~\cite{dampe2019measurement}. For the gamma rate estimation, we assume the Crab Nebula spectrum follows a modified log-parabola function, taken from~\cite{ALEKSIC201530}. Details of the simulation parameters are summarized in Table~\ref{tab:simulation_params}.

Protons are the dominant source of background and are indistinguishable from gamma rays at the trigger level. Considering a DAQ bandwidth of about 50~kHz, a minimum triggering threshold of 20~p.e. can be set. At low trigger thresholds, the NSB dominates the rates and can rapidly saturate the data acquisition system. Therefore, the main goal of the trigger is to separate proton/gamma showers from NSB.

\section{Trigger logic}

\begin{wrapfigure}{r}{0.65\textwidth}
    \centering
    \includegraphics[angle=270, width=10.0cm]{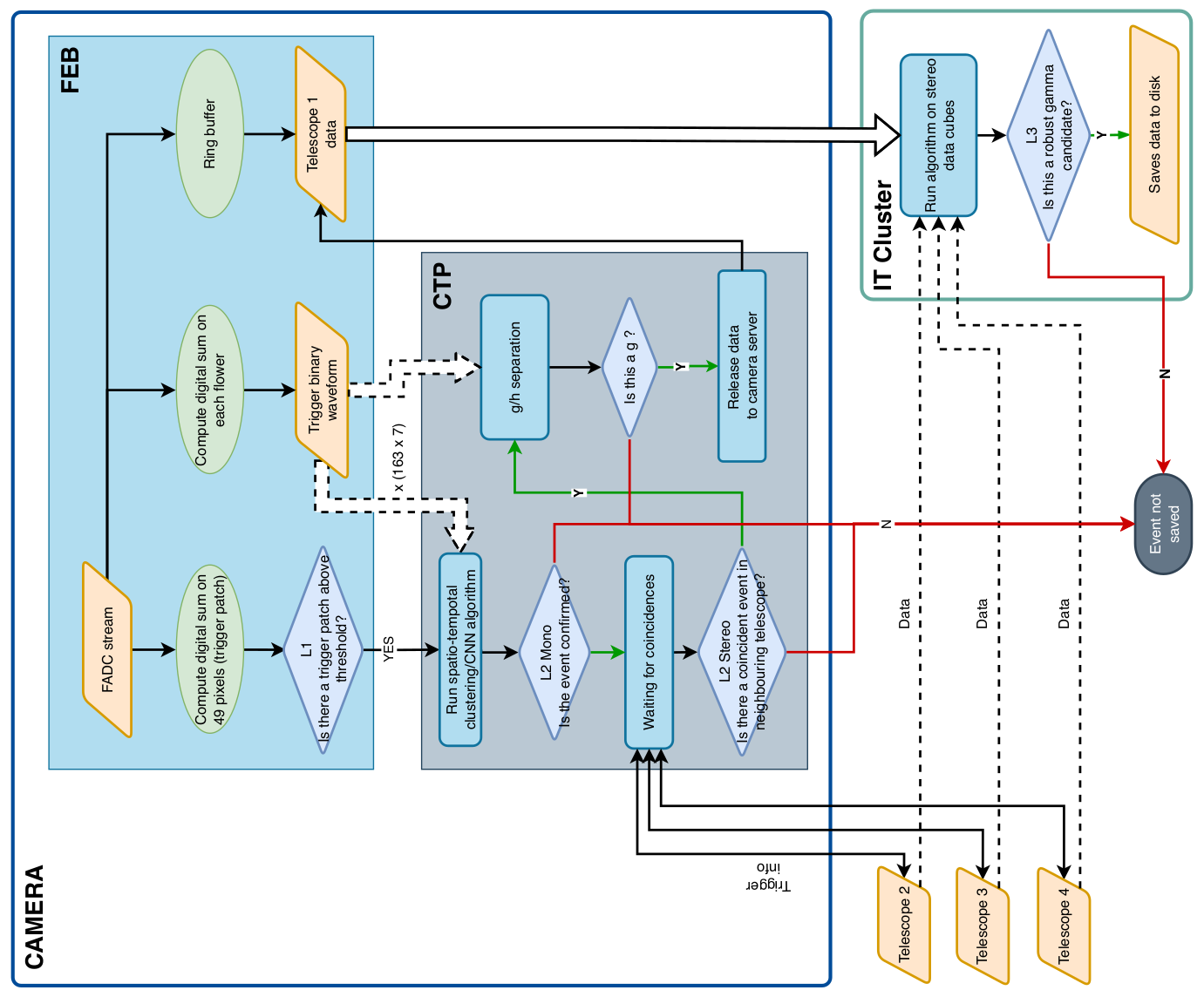}
    \caption{Trigger Flow Diagram.}
    \label{fig:TriggerFlow}
\end{wrapfigure}

The fully digital readout of AdvCam enables a multi-level trigger hierarchy that allows a reduction of the detection energy threshold, while keeping a manageable rate for the data written to disk. This hierarchy splits into a low-level logic, hardware-based, including a Level~1 and a Level~2 Trigger, and a high-level logic, software-based, corresponding to the Level~3 Trigger (Figure~\ref{fig:TriggerFlow}).

\textbf{Level~1 Trigger.} The initial trigger stage, dubbed L1, is a digital sum computed at the Front-End-Boards (FEBs) in a real-time, 1~GHz, rate, on a group of 49 pixels, referred to, later in the text, as L1 trigger patches or super flower (Figure~\ref{fig:PMT_vs_SiPM}). L1 trigger patches overlaps, so that they are computed centered on each 7-pixel flower, leaving no dead or inefficient regions in the camera. A simple threshold cut is applied to the super flower digital sum, so that for each 7-pixel flower a L1~trigger signal is produced. A different use of the L1~signals will be done, depending on the chosen algorithm for the Level 2 Trigger.



\textbf{Level~2 Trigger.} The Level 2 trigger, dubbed L2, is performed at the Central Trigger Processor (CTP), a camera electronics subsystem involving a set of Field-Programmable Gate Arrays (FPGA), based on the information received at the CTP from the FEBs. The L2 trigger serves as a {\itshape Camera Trigger}, so that the digitized pixel content, stored in the FEB buffers, is read-out and sent to the DAQ system at the Counting House. The main purpose of the L2 trigger is to further reduce the camera’s DAQ rate based on the output of the L1 trigger. This {\itshape Camera Trigger} is based on 2 steps, the so-called Local Level 2 Trigger and Topo-Stereo trigger Level 2 Trigger~\cite{L_pez_Coto_2016}.

\textbf{Local L2 Trigger.} The specific algorithm for generating the Local L2 trigger signal is still under discussion. Two main approaches are considered: an unsupervised clustering algorithm, namely the Density-Based Spatial Clustering of Applications with Noise (DBSCAN) algorithm~\cite{ester1996density}, and another option based on Convolution Neural Networks (CNN)~\cite{lecun2002gradient}.

\begin{figure}
    \centering
    \includegraphics[height=3.2cm]{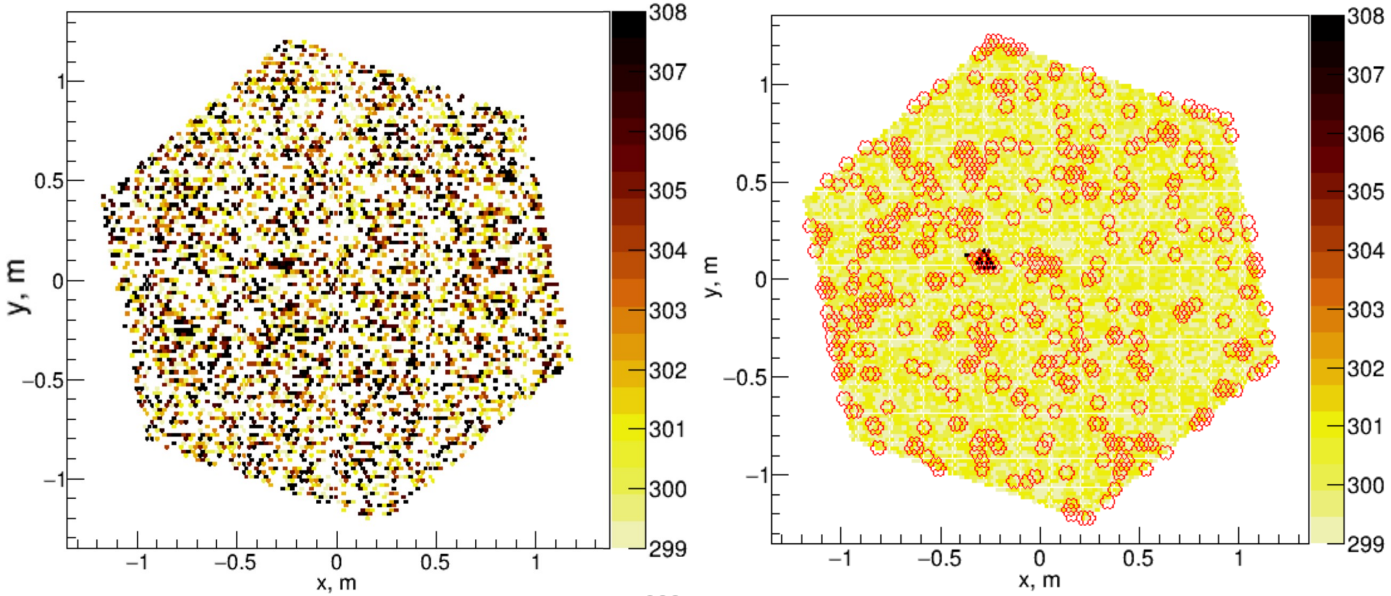}
    \includegraphics[height=3.2cm]{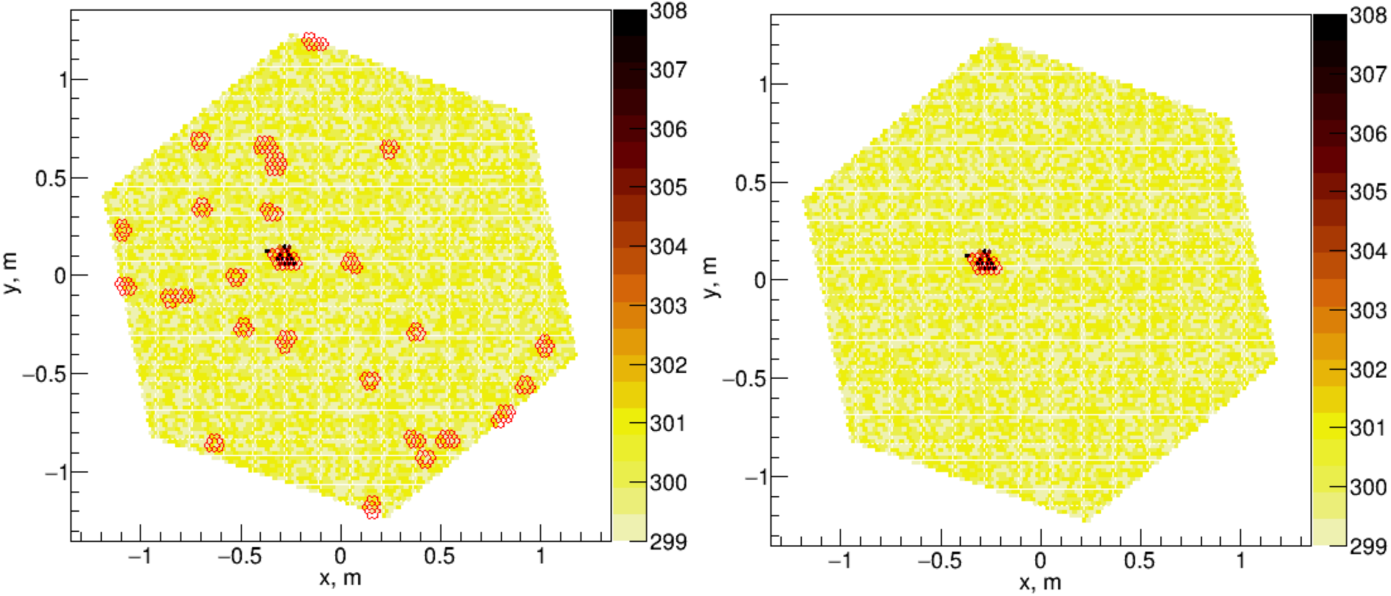}
    \caption{The 12~GeV on-axis gamma ray with a core located 180~m from the telescope. We plot the instantaneous ADC value of the digitized waveform by selecting the time frame with the maximum signal. (\textbf{First left}): The nominal NSB rate is 268~MHz/channel, and the nominal electronic noise 3.94 ADC counts RMS. (\textbf{Second left}): The NSB rate is set to 0, and the electronic noise is set to 0.1 RMS ADC counts. The fine red lines indicate the contours of the flower-like pattern where the digital sum exceeds the threshold, superimposed over the entire readout interval. (\textbf{First right}): Only spatially coincident flowers are shown. (\textbf{Second right}): Only flowers that are spatially and temporally coincident are shown. 
    Note: simulations with a 268~MHz/channel NSB rate and an electronic noise level of 3.94~ADC are used in the analysis, with visualizations superimposed on noise-free waveforms.}
    \label{fig:DBSCAN}
\end{figure}

\textbf{Local L2 Trigger: DBSCAN.} The DBSCAN algorithm is a natural choice for distinguishing signal from background when searching for slight increases in photon density in part of the camera relative to the NSB. It efficiently identifies dense clusters of arbitrary shape in a multidimensional space. The algorithm requires two input parameters, \textit{Eps, MinPts}. For the distance metric, we chose the Euclidean metric. The \textit{Eps} and \textit{MinPts} define least dense cluster in the data: \textit{Eps}-Neighborhood contains at least \textit{MinPts} points. Both of them can be defined manually, but the approximate \textit{Eps} value is often determined using a sorted k-distance graph~\cite{ester1996density}. This method can be routinely used for image cleaning, however applying it to the time sequence, as we record waveforms, requires scaling the time dimension relative to the two spatial axes representing the channel positions in the camera. Figure~\ref{fig:DBSCAN} illustrates DBSCAN’s ability to analyze waveforms, successfully detecting a 12~GeV gamma ray.

In the context of AdvCam, DBSCAN will be fed with the L1 signals from the 1171 7-pixel flowers. It is currently considered as the primary option for the Local L2 Trigger, as DBSACAN-like algorithms can run in the CTP FPGAs to cope with the 1~GHz input rate, thus requiring no L1 filter in the FEBs. Additionally, DBSCAN can provide the  shower position in the camera, to be later used for the Topo-Stereo L2 trigger. 

\begin{figure}
    \centering
    \includegraphics[width=15.0cm]{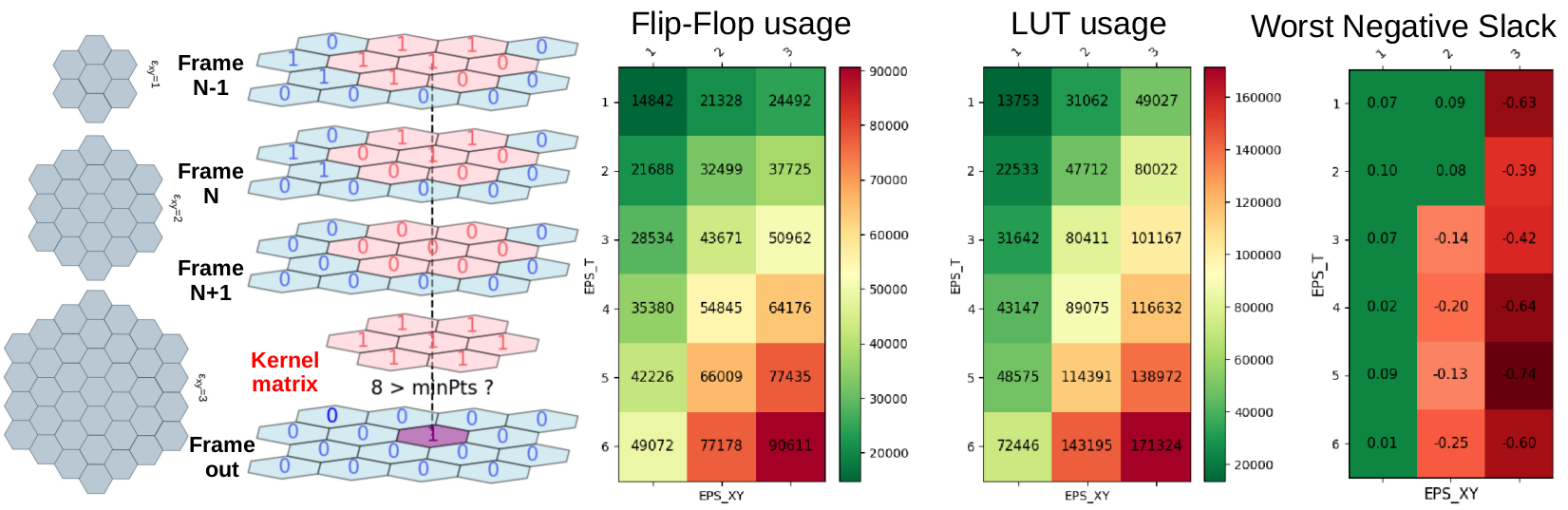}
    \caption{TDSCAN algorithm and FPGA implementation. (\textbf{Left}): Kernel sizes used for convolution. (\textbf{Center}): Hexagonal 3D~convolution over two spatial coordinates and one temporal dimension (frames N-1, N, N+1 ...), producing the output frame. (\textbf{Right}): Resource usage and timing at 350~MHz target frequency.}
    \label{fig:TDSCAN}
\end{figure}

\textbf{Local L2 Trigger: TDSCAN.}  We propose an FPGA-efficient approximation of DBSCAN that uses 3D convolution with a hexagonal kernel. TDSCAN stand for Trigger Distributed Spatial Convolution Accelerator Network (Figure~\ref{fig:TDSCAN}). It has three hyperparameters: \textit{Eps\textsubscript{t}}, \textit{Eps\textsubscript{XY}}, and \textit{MinPts}, where \textit{Eps} is defined separately for the temporal and spatial dimensions. The algorithm does not differentiate between different clusters and has a fixed latency for a given \textit{Eps\textsubscript{t}}, \textit{Eps\textsubscript{XY}}, making it suitable for working with data streams. The TDSCAN algorithm works in simulation and has been implemented on the Xilinx KCU105 board. It can run at a frequency of \textbf{350~MHz} with a latency of approximately \textbf{14.28~ns} (\textit{Eps\textsubscript{t}}=1, \textit{Eps\textsubscript{XY}}=1), and 1~GHz version in development.

\textbf{Local L2 Trigger:  Convolutional Neural Networks.} CNN is a special type of Deep Neural Networks, especially effective at handling 2D images, as demonstrated by the \texttt{CTLearn}~\cite{miener_2025_15065761} package for IACT offline reconstruction~\cite{Miener:2021ixs}. These networks consist of convolutional layers combined with pooling layers. The convolutional layers apply filters that extract features (like edges, shapes, and textures). In contrast, the pooling layers perform sub-sampling to maintain key features and spatial relationships between pixels, thus enhancing the network’s efficiency and minimizing memory requirements. Finally, fully connected layers provide high-level reasoning based on the extracted feature maps, leading to the final prediction.

CNN algorithms can process full-camera raw waveforms, of 5–10~ns long, to distinguish NSB from showers, or gammas from hadrons. They also provide shower position candidates for the Topo-Stereo L2 trigger. Once optimized for CTP FPGAs, they are a viable option for the Local L2 Trigger algorithm. However, the expected maximum inference rate for CNN L2 Trigger algorithms is about 1~MHz. This requires using an L1 trigger filter at the FEB level to reduce the rate from ~1 GHz to ~1 MHz, based on a simple digital sum cut.

\begin{figure}
    \centering
    \includegraphics[height=4.0cm]{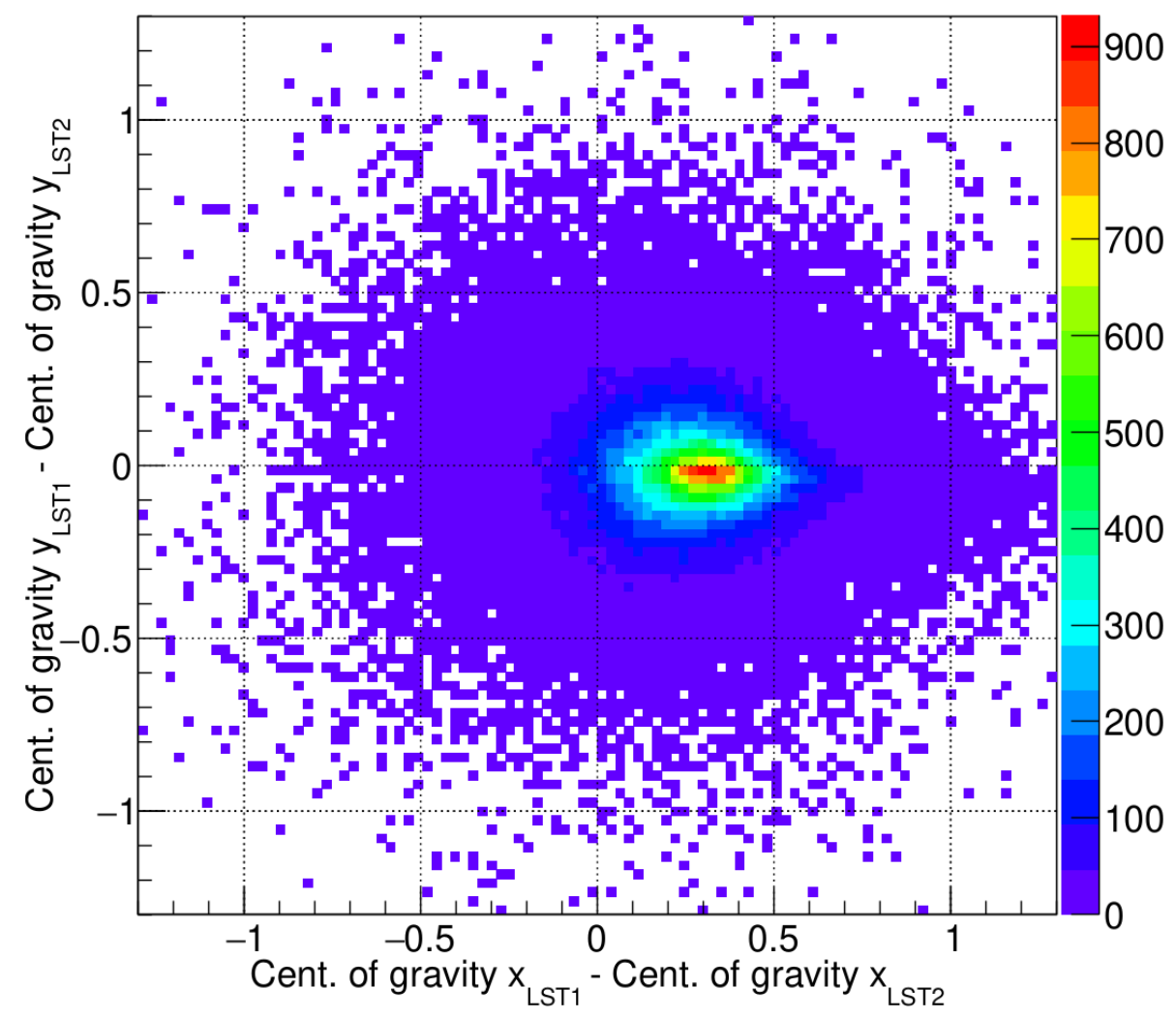}
    \includegraphics[height=4.0cm]{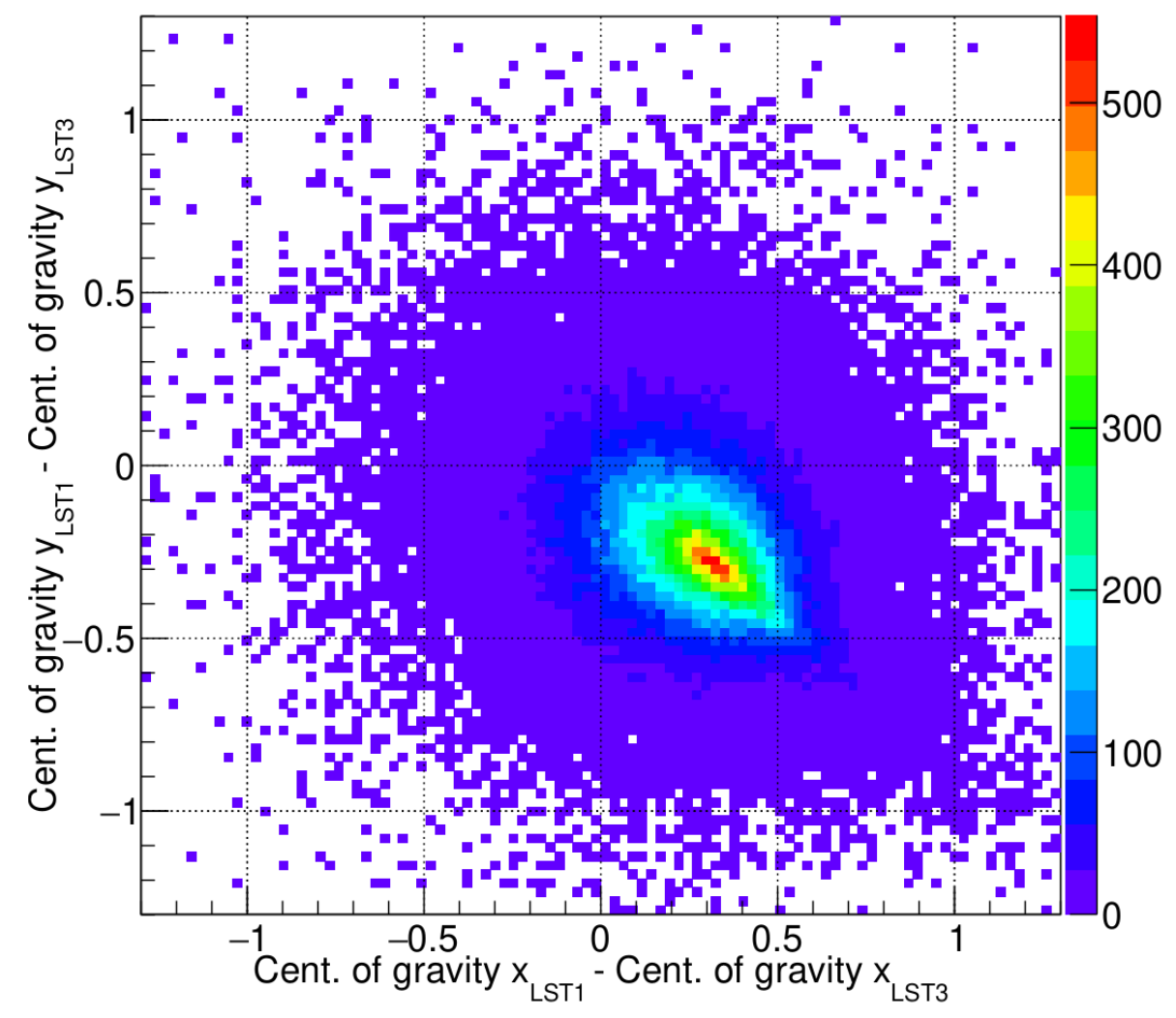}
    \includegraphics[height=4.0cm]{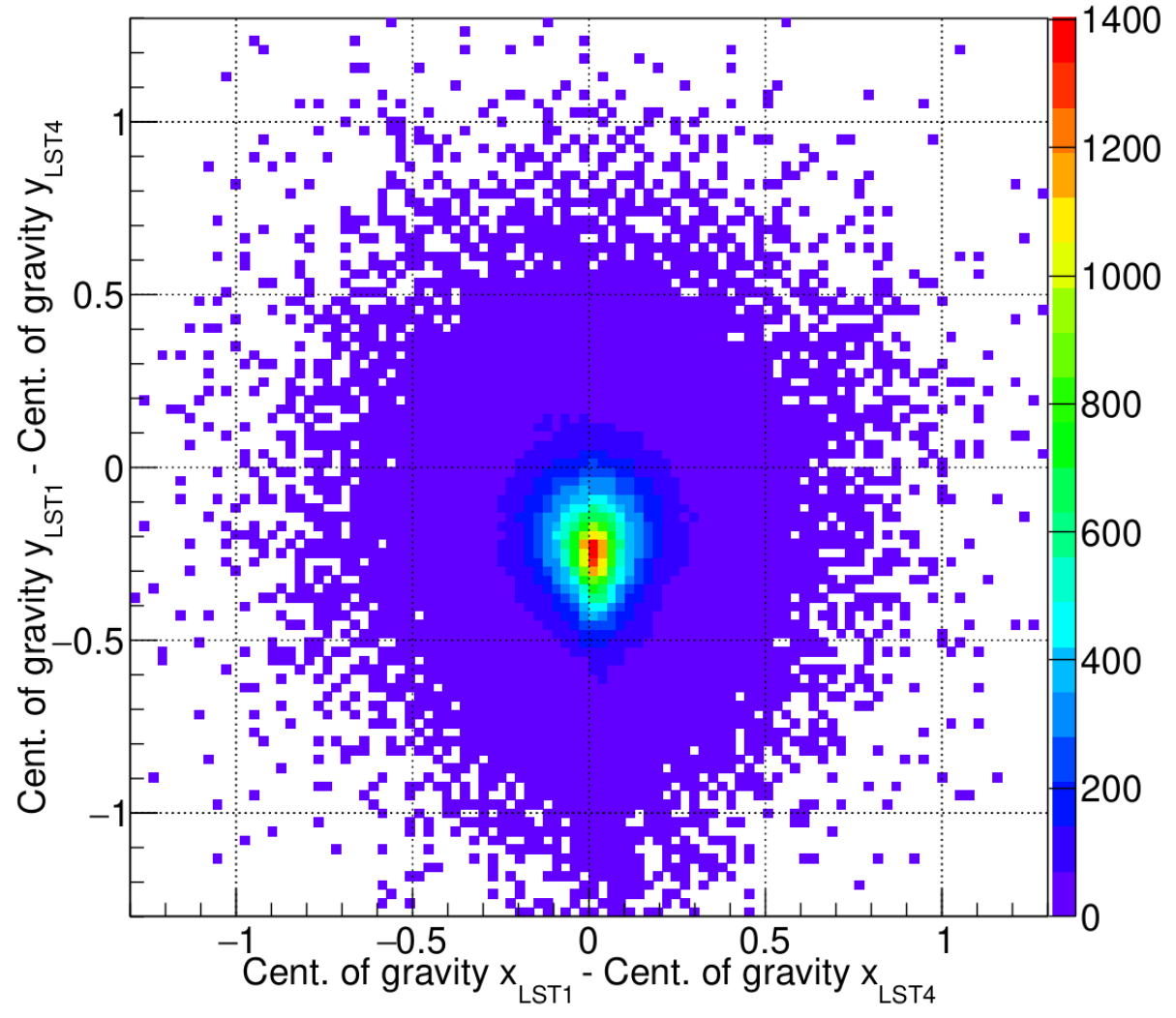}
    \caption{ The differences in camera signal barycenters for proton events between LST-1 and other LSTs: LST-2 (left), LST-3 (center), and LST-4 (right).}
    \label{fig:LSTtopo}
\end{figure}

\textbf{Topo-Stereo L2 Trigger.} Once an AdvCam has produced a Local L2 Trigger signal it is shared with the rest of the cameras of the LST array, so that at each camera CTP a Topo-Stereo L2 trigger is evaluated. The corresponding Stereo L2 trigger is thus considered the {\itshape Camera Trigger}, providing the desired data acquisition system (DAQ) rate of $\sim$40-50~kHz. Figure~\ref{fig:LSTtopo} illustrates the topological trigger principle: knowing the shower position in one telescope allows us to predict its location in the other three telescopes. This effectively reduces false stereo coincidences.

\begin{figure}
    \centering
    \includegraphics[height=3.5cm]{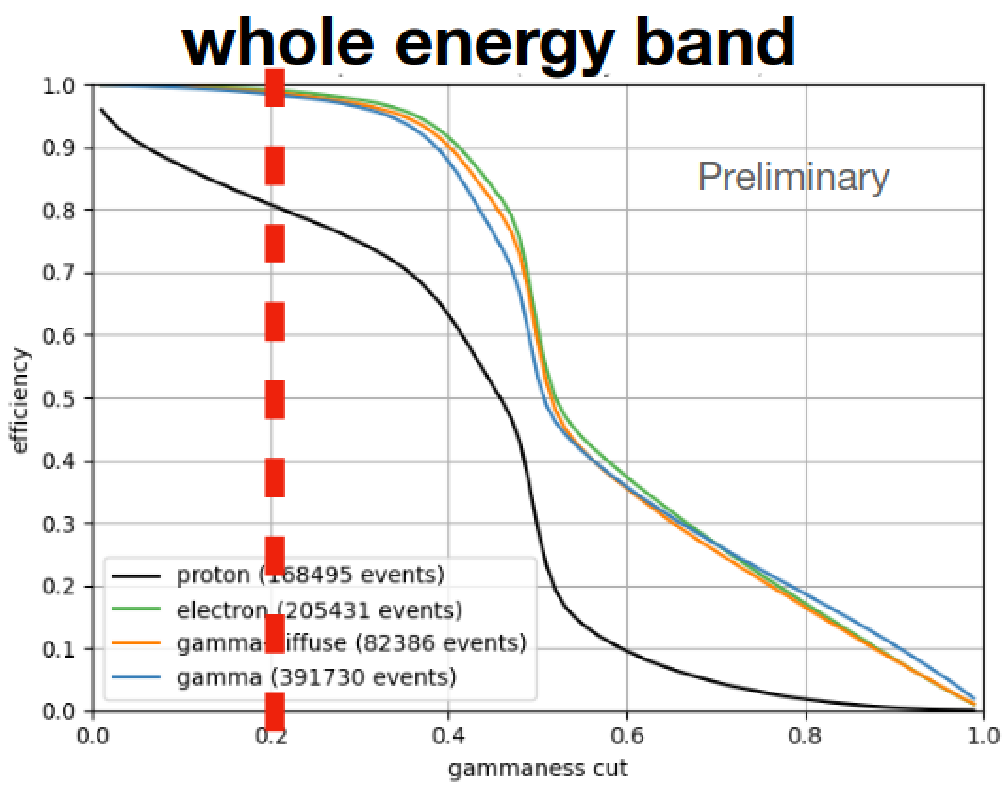}
    \includegraphics[height=3.5cm]{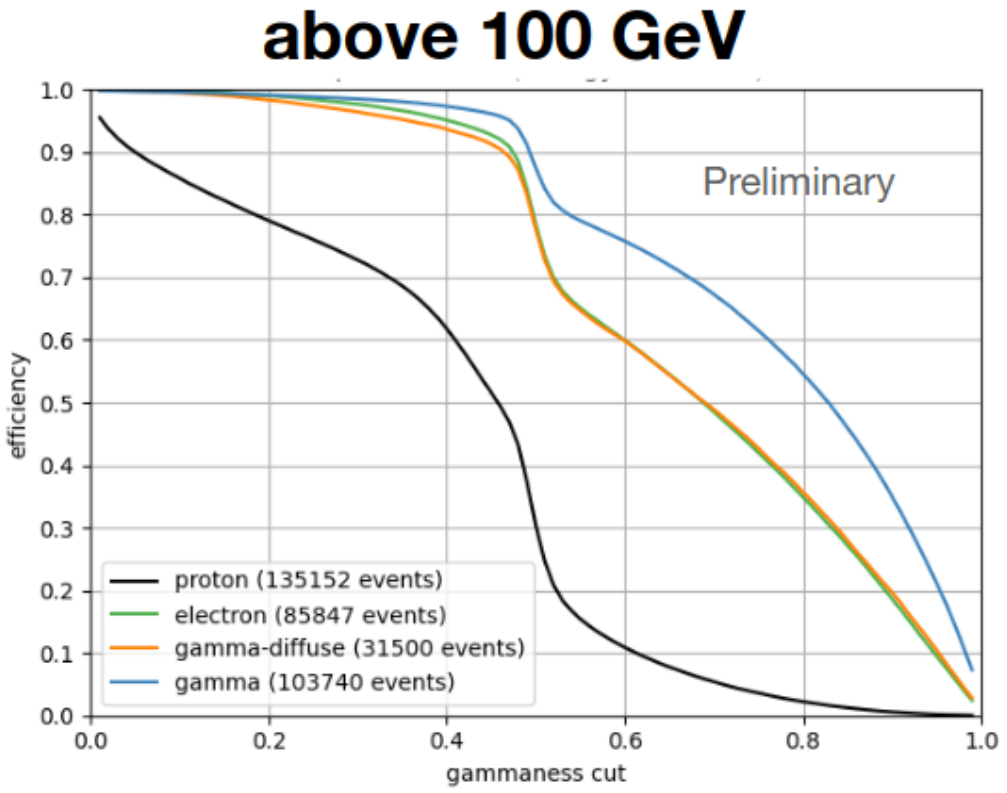}
    \includegraphics[height=3.5cm]{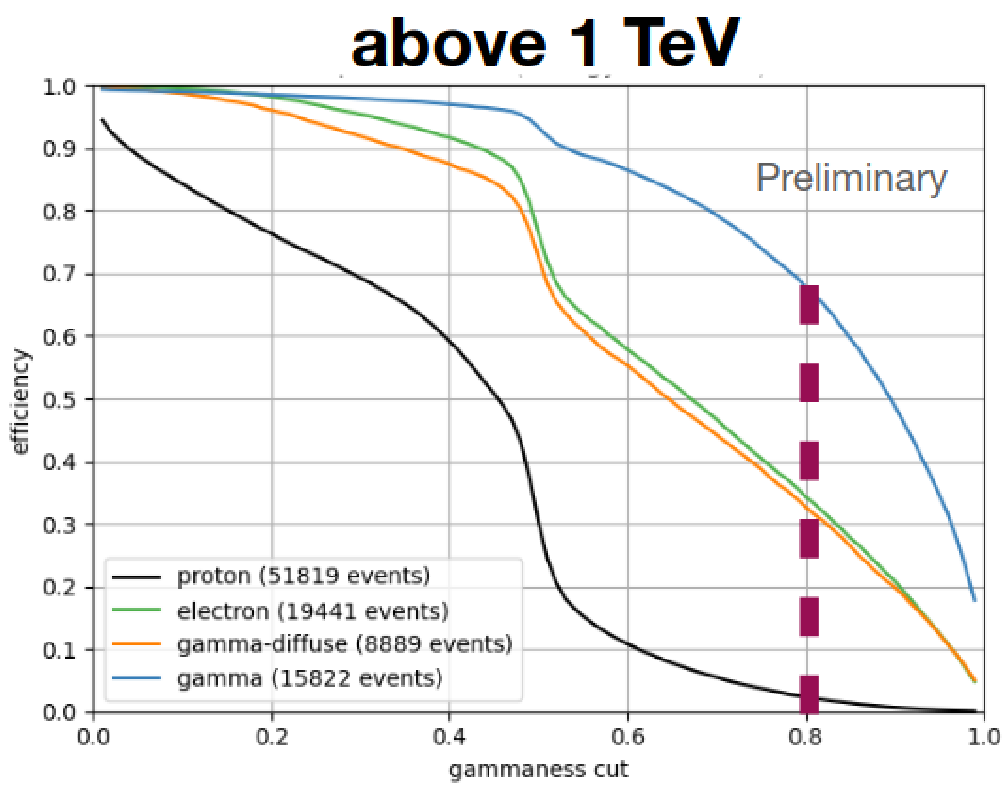}
    \caption{Efficiency curves for on-axis gamma rays, diffuse gamma rays, electrons, and protons as a function of the classifier output (gammaness). The three panels correspond to different energy ranges: whole energy band (\textbf{left}), above 100 GeV (\textbf{middle}), and above 1 TeV (\textbf{right}).}
    \label{fig:topolia2}
\end{figure}

\textbf{Level 3 trigger} The L2 trigger logic ensures that a minimum of two AdvCams are read out for each detected event. Following this, a Level-3 (L3) software-based trigger is applied within the DAQ system, enabling the real-time processing of raw waveforms from all read-out cameras. We use \texttt{CTLearn}, a high-level IACT data analysis package, to implement a simple CNN (\textasciitilde 200k model parameters) with three convolutional layers (32, 32, 64 filters) and a dense head (256, 64 neurons) for binary classification. By porting a comparable \texttt{CTLearn} model to the OpenVINO framework, we achieved real-time inference at \textasciitilde 40k frames per second on an Intel Arria 10 PAC FPGA, matching the desired DAQ rate. The applied floating point (FP) precision reduction from FP32 to FP16 preserved the physics performance at $\mathcal{O} \big(10^{-3}\big)$. Further precision reductions (FP12, INT9, INT8) are under study and also the usage of quantization-aware training via \texttt{hls4ml}~\cite{Duarte_2018} to minimize a potential performance drop is planned.

This L3 software-based trigger approach offers the ability of real-time suppressing of proton-like cosmic-ray events. Cosmic rays are the dominant background occurring up to $\sim10^5 $ times more frequently than gamma rays (depending on observational conditions). Thereby, this approach can reduce data storage and computational demands downstream. In the left panel of Fig.~\ref{fig:topolia2}, the red vertical line indicates a very loose gammaness cut that preserves nearly all gamma rays and electrons while reducing data volume by 20\% (rejection of protons), demonstrating the model’s utility for online triggering. In the right panel of Fig.~\ref{fig:topolia2}, the purple line marks a cut that retains \textasciitilde 70\% of on-axis gammas above 1 TeV, a typical benchmark for optimizing high-level off-line analysis selections. Notably, the model effectively rejects nearly all protons above 1 TeV while maintaining an on-axis gamma-ray efficiency of \textasciitilde 70\%, highlighting its strong background suppression capabilities in the high-energy regime. Across the whole energy range, this cut rejects more than half of the simulated proton events.

\section{Results and conclusions}
\begin{figure}
    \centering
    \begin{overpic}[height=5.0cm]{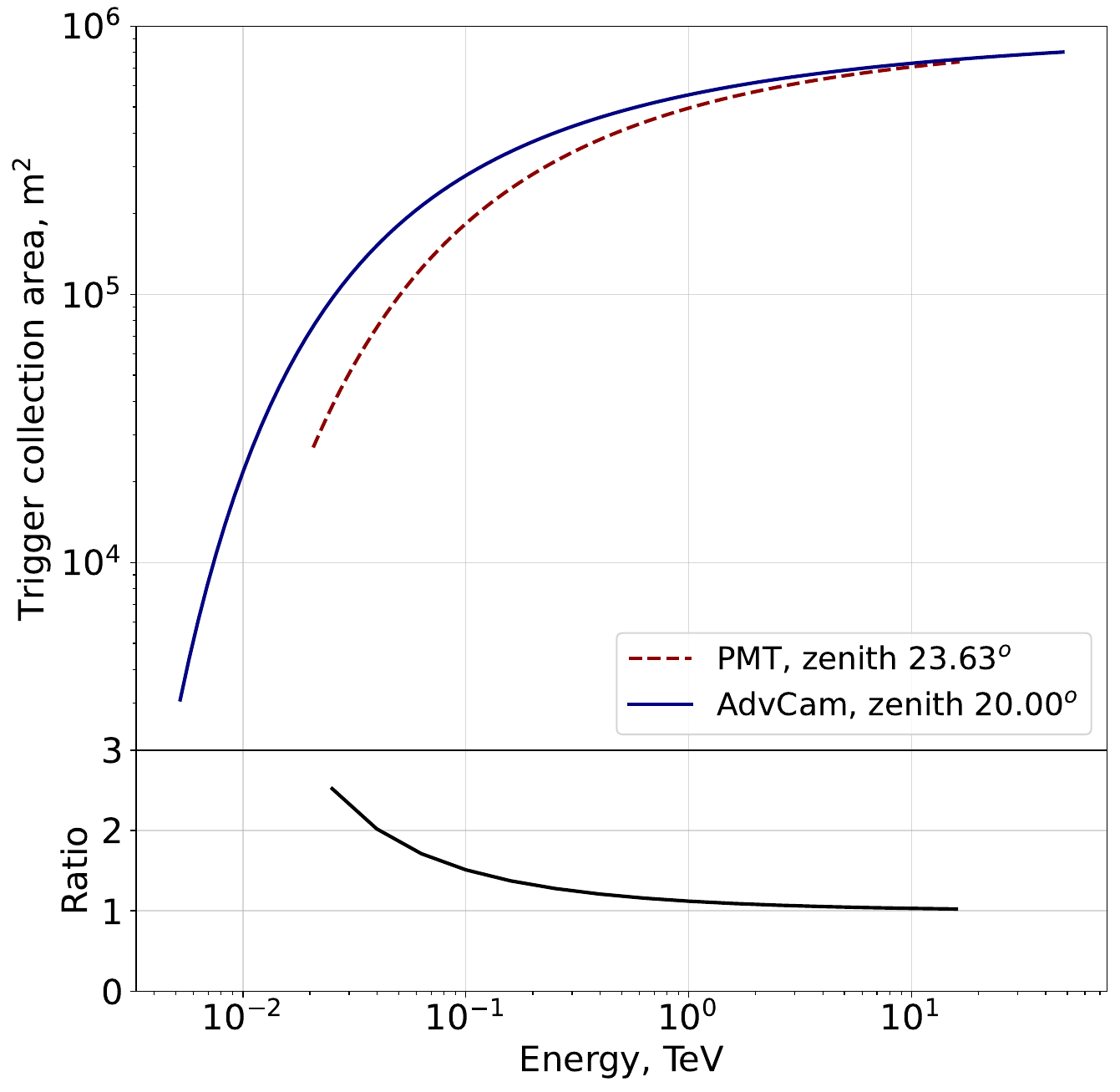} 
        \put(50,60){\textcolor{gray}{Preliminary}}
    \end{overpic}
    \begin{overpic}[height=5.0cm]{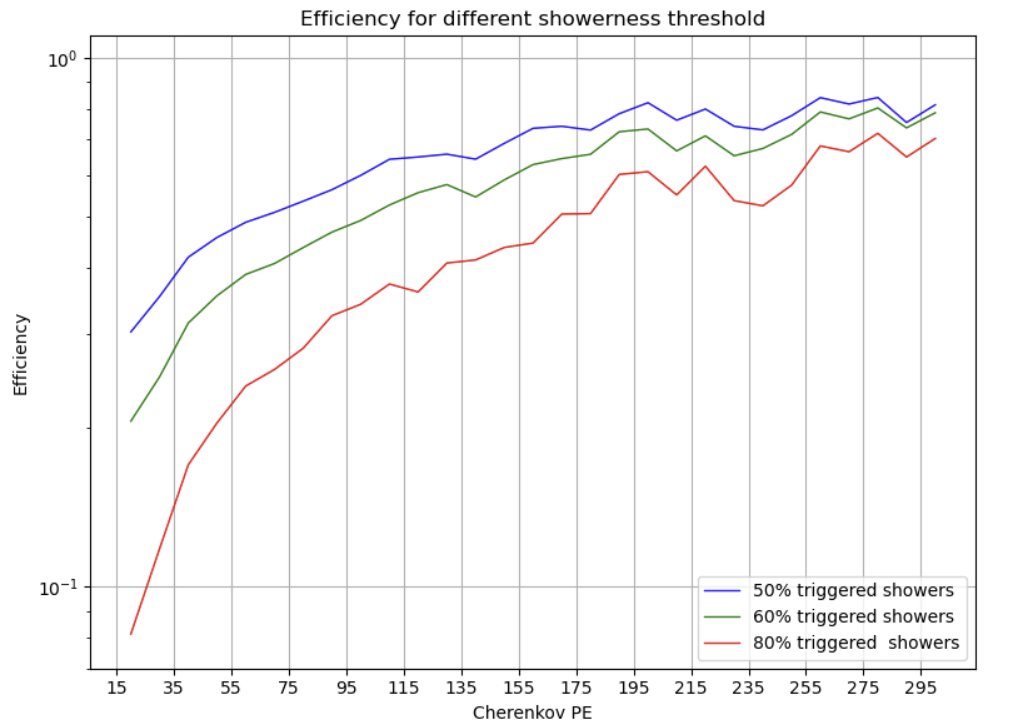} 
        \put(55,30){\textcolor{gray}{Preliminary}}
    \end{overpic}
    \caption{(\textbf{Left}): (\textbf{Preliminary}) Trigger collection area as a function of gamma energy for the current LST-1 and the AdvCam with the newly proposed trigger scheme and corresponding ratio between the two configurations. (\textbf{Right}): Trigger efficiency as a function of the number of p.e. per event for the single CNN model, applying 3 different cuts in the predicted showerness value.}
    \label{fig:trg_effArea_AdvCam_vs_PMT}
\end{figure}

We present a comparison of the trigger collection area as a function of gamma-ray energy for the current LST-1~\cite{Abe_2023} and a simulation of the AdvCam using the newly proposed trigger scheme on Figure~\ref{fig:trg_effArea_AdvCam_vs_PMT} (left).  As expected, the enhancement is more significant in the low-energy range, which is primarily the responsibility of LSTs. The AdvCam, with its novel readout system, can significantly enhance the triggering capabilities of a single LST. The DBSCAN algorithm has great potential not only for providing a sophisticated trigger but also for cleaning the data cube.

We present the trigger efficiency of the Local L2 CNN-based trigger scheme as a function of the number of detected p.e. The CNNs have been trained on simulated camera {\itshape patches}, i.e., portions of the camera with 7 superflowers (343 pixels), containing either NSB-induced or NSB and shower-induced p.e. {\itshape Shower-ness} refers to the classifier values returned by the trained CNN for distinguishing between shower and NSB events. For each camera patch, a {\itshape shower-ness} score is computed, and a full event trigger is determined by applying a logical OR operation across all predictions. Figure~\ref{fig:trg_effArea_AdvCam_vs_PMT} (right) displays the results for three different thresholds applied to the shower-ness value predicted by the CNN. These thresholds correspond to different levels of stringency, defined by the percentage of triggered events that are actual showers (i.e., the ratio of shower triggers to total triggers). Despite its simplicity, the single CNN model demonstrates the ability to effectively distinguish between patches dominated by NSB and those containing real shower signals. Notably, the trigger shows strong performance even at low p.e. counts. In addition, the use of hexagonal convolution, the lightweight architecture, and ongoing efforts toward model size reduction make this CNN trigger scheme a promising candidate for implementation on FPGAs.

\footnotesize
\bibliographystyle{unsrturl}
\bibliography{biblio.bib}

\textbf{Full Author List: CTAO-LST Project}

\tiny{\noindent
K.~Abe$^{1}$,
S.~Abe$^{2}$,
A.~Abhishek$^{3}$,
F.~Acero$^{4,5}$,
A.~Aguasca-Cabot$^{6}$,
I.~Agudo$^{7}$,
C.~Alispach$^{8}$,
D.~Ambrosino$^{9}$,
F.~Ambrosino$^{10}$,
L.~A.~Antonelli$^{10}$
% ...
The CTAO-LST Project
}\\

\tiny{\noindent$^{1}${Department of Physics, Tokai University, 4-1-1, Kita-Kaname, Hiratsuka, Kanagawa 259-1292, Japan}.
$^{2}${Institute for Cosmic Ray Research, University of Tokyo, 5-1-5, Kashiwa-no-ha, Kashiwa, Chiba 277-8582, Japan}.
$^{3}${INFN and Università degli Studi di Siena, Dipartimento di Scienze Fisiche, della Terra e dell'Ambiente (DSFTA), Sezione di Fisica, Via Roma 56, 53100 Siena, Italy}.
$^{4}${Université Paris-Saclay, Université Paris Cité, CEA, CNRS, AIM, F-91191 Gif-sur-Yvette Cedex, France}.
$^{5}${FSLAC IRL 2009, CNRS/IAC, La Laguna, Tenerife, Spain}.
$^{6}${Departament de Física Quàntica i Astrofísica, Institut de Ciències del Cosmos, Universitat de Barcelona, IEEC-UB, Martí i Franquès, 1, 08028, Barcelona, Spain}.
$^{7}${Instituto de Astrofísica de Andalucía-CSIC, Glorieta de la Astronomía s/n, 18008, Granada, Spain}.
$^{8}${Department of Astronomy, University of Geneva, Chemin d'Ecogia 16, CH-1290 Versoix, Switzerland}.
$^{9}${INFN Sezione di Napoli, Via Cintia, ed. G, 80126 Napoli, Italy}.
$^{10}${INAF - Osservatorio Astronomico di Roma, Via di Frascati 33, 00040, Monteporzio Catone, Italy}.
% ...
}

\acknowledgments 
\tiny{
We gratefully acknowledge financial support from the following agencies and organisations:
Conselho Nacional de Desenvolvimento Cient\'{\i}fico e Tecnol\'{o}gico (CNPq), Funda\c{c}\~{a}o de Amparo \`{a} Pesquisa do Estado do Rio de Janeiro (FAPERJ), Funda\c{c}\~{a}o de Amparo \`{a} Pesquisa do Estado de S\~{a}o Paulo (FAPESP), Funda\c{c}\~{a}o de Apoio \`{a} Ci\^encia, Tecnologia e Inova\c{c}\~{a}o do Paran\'a - Funda\c{c}\~{a}o Arauc\'aria, Ministry of Science, Technology, Innovations and Communications (MCTIC), Brasil;
Ministry of Education and Science, National RI Roadmap Project DO1-153/28.08.2018, Bulgaria;
Croatian Science Foundation (HrZZ) Project IP-2022-10-4595, Rudjer Boskovic Institute, University of Osijek, University of Rijeka, University of Split, Faculty of Electrical Engineering, Mechanical Engineering and Naval Architecture, University of Zagreb, Faculty of Electrical Engineering and Computing, Croatia;
Ministry of Education, Youth and Sports, MEYS  LM2023047, EU/MEYS CZ.02.1.01/0.0/0.0/16\_013/0001403, CZ.02.1.01/0.0/0.0/18\_046/0016007, CZ.02.1.01/0.0/0.0/16\_019/0000754, CZ.02.01.01/00/22\_008/0004632 and CZ.02.01.01/00/23\_015/0008197 Czech Republic;
CNRS-IN2P3, the French Programme d’investissements d’avenir and the Enigmass Labex, 
This work has been done thanks to the facilities offered by the Univ. Savoie Mont Blanc - CNRS/IN2P3 MUST computing center, France;
Max Planck Society, German Bundesministerium f{\"u}r Bildung und Forschung (Verbundforschung / ErUM), Deutsche Forschungsgemeinschaft (SFBs 876 and 1491), Germany;
Istituto Nazionale di Astrofisica (INAF), Istituto Nazionale di Fisica Nucleare (INFN), Italian Ministry for University and Research (MUR), and the financial support from the European Union -- Next Generation EU under the project IR0000012 - CTA+ (CUP C53C22000430006), announcement N.3264 on 28/12/2021: ``Rafforzamento e creazione di IR nell’ambito del Piano Nazionale di Ripresa e Resilienza (PNRR)'';
ICRR, University of Tokyo, JSPS, MEXT, Japan;
JST SPRING - JPMJSP2108;
Narodowe Centrum Nauki, grant number 2023/50/A/ST9/00254, Poland;
The Spanish groups acknowledge the Spanish Ministry of Science and Innovation and the Spanish Research State Agency (AEI) through the government budget lines
PGE2022/28.06.000X.711.04,
28.06.000X.411.01 and 28.06.000X.711.04 of PGE 2023, 2024 and 2025,
and grants PID2019-104114RB-C31,  PID2019-107847RB-C44, PID2019-104114RB-C32, PID2019-105510GB-C31, PID2019-104114RB-C33, PID2019-107847RB-C43, PID2019-107847RB-C42, PID2019-107988GB-C22, PID2021-124581OB-I00, PID2021-125331NB-I00, PID2022-136828NB-C41, PID2022-137810NB-C22, PID2022-138172NB-C41, PID2022-138172NB-C42, PID2022-138172NB-C43, PID2022-139117NB-C41, PID2022-139117NB-C42, PID2022-139117NB-C43, PID2022-139117NB-C44, PID2022-136828NB-C42, PDC2023-145839-I00 funded by the Spanish MCIN/AEI/10.13039/501100011033 and “and by ERDF/EU and NextGenerationEU PRTR; the "Centro de Excelencia Severo Ochoa" program through grants no. CEX2019-000920-S, CEX2020-001007-S, CEX2021-001131-S; the "Unidad de Excelencia Mar\'ia de Maeztu" program through grants no. CEX2019-000918-M, CEX2020-001058-M; the "Ram\'on y Cajal" program through grants RYC2021-032991-I  funded by MICIN/AEI/10.13039/501100011033 and the European Union “NextGenerationEU”/PRTR and RYC2020-028639-I; the "Juan de la Cierva-Incorporaci\'on" program through grant no. IJC2019-040315-I and "Juan de la Cierva-formaci\'on"' through grant JDC2022-049705-I. They also acknowledge the "Atracci\'on de Talento" program of Comunidad de Madrid through grant no. 2019-T2/TIC-12900; the project "Tecnolog\'ias avanzadas para la exploraci\'on del universo y sus componentes" (PR47/21 TAU), funded by Comunidad de Madrid, by the Recovery, Transformation and Resilience Plan from the Spanish State, and by NextGenerationEU from the European Union through the Recovery and Resilience Facility; “MAD4SPACE: Desarrollo de tecnolog\'ias habilitadoras para estudios del espacio en la Comunidad de Madrid" (TEC-2024/TEC-182) project funded by Comunidad de Madrid; the La Caixa Banking Foundation, grant no. LCF/BQ/PI21/11830030; Junta de Andaluc\'ia under Plan Complementario de I+D+I (Ref. AST22\_0001) and Plan Andaluz de Investigaci\'on, Desarrollo e Innovaci\'on as research group FQM-322; Project ref. AST22\_00001\_9 with funding from NextGenerationEU funds; the “Ministerio de Ciencia, Innovaci\'on y Universidades”  and its “Plan de Recuperaci\'on, Transformaci\'on y Resiliencia”; “Consejer\'ia de Universidad, Investigaci\'on e Innovaci\'on” of the regional government of Andaluc\'ia and “Consejo Superior de Investigaciones Cient\'ificas”, Grant CNS2023-144504 funded by MICIU/AEI/10.13039/501100011033 and by the European Union NextGenerationEU/PRTR,  the European Union's Recovery and Resilience Facility-Next Generation, in the framework of the General Invitation of the Spanish Government’s public business entity Red.es to participate in talent attraction and retention programmes within Investment 4 of Component 19 of the Recovery, Transformation and Resilience Plan; Junta de Andaluc\'{\i}a under Plan Complementario de I+D+I (Ref. AST22\_00001), Plan Andaluz de Investigaci\'on, Desarrollo e Innovación (Ref. FQM-322). ``Programa Operativo de Crecimiento Inteligente" FEDER 2014-2020 (Ref.~ESFRI-2017-IAC-12), Ministerio de Ciencia e Innovaci\'on, 15\% co-financed by Consejer\'ia de Econom\'ia, Industria, Comercio y Conocimiento del Gobierno de Canarias; the "CERCA" program and the grants 2021SGR00426 and 2021SGR00679, all funded by the Generalitat de Catalunya; and the European Union's NextGenerationEU (PRTR-C17.I1). This research used the computing and storage resources provided by the Port d’Informaci\'o Cient\'ifica (PIC) data center.
State Secretariat for Education, Research and Innovation (SERI) and Swiss National Science Foundation (SNSF), Switzerland;
The research leading to these results has received funding from the European Union's Seventh Framework Programme (FP7/2007-2013) under grant agreements No~262053 and No~317446;
This project is receiving funding from the European Union's Horizon 2020 research and innovation programs under agreement No~676134;
ESCAPE - The European Science Cluster of Astronomy \& Particle Physics ESFRI Research Infrastructures has received funding from the European Union’s Horizon 2020 research and innovation programme under Grant Agreement no. 824064.}

\end{document}